# Cold Creep of Titanium: Analysis of stress relaxation using synchrotron diffraction and crystal plasticity simulations

*Yi Xiong* [a*]*, Phani Karamched* [a]*, Chi-Toan Nguyen* [b]*, David M Collins* [c]*, Christopher M Magazzeni* [a]*, Edmund Tarleton* [d,a]*, Angus J Wilkinson* [a]

[a] *Department of Materials, University of Oxford, Parks Road, Oxford, OX1 3PH, United Kingdom*

[b] *Safran SA, Safran Tech, Department of Materials and Process, Rue des Jeunes Bois – Châteaufort, 78772 Many-les-Hameaux, France*

[c] *School of Metallurgy and Materials, University of Birmingham, Edgbaston, Birmingham, B15 2TT, United Kingdom*

[d] *Department of Engineering Science, University of Oxford, Parks Road, Oxford, OX1 3PJ, United Kingdom*

---

[*] Corresponding author

E-mail address: yi.xiong@materials.ox.ac.uk (Yi Xiong)

# Abstract

It is well known that titanium and some titanium alloys creep at ambient temperature, resulting in a significant fatigue life reduction when a stress dwell is included in the fatigue cycle. It is thought that localised *time dependent plasticity* in 'soft' grains oriented for easy plastic slip leads to load shedding and an increase in stress within a neighbouring 'hard' grain that is poorly oriented for easy slip. Quantifying this time dependent plasticity process is key to successfully predicting the complex cold dwell fatigue problem. In this work, synchrotron X-ray diffraction during stress relaxation experiments was performed to characterise the time dependent plastic behaviour of commercially pure titanium (grade 4). Lattice strains were measured by tracking the diffraction peak shift from multiple plane families (21 diffraction rings) as a function of their orientation with respect to the loading direction. The critical resolved shear stress, activation energy and activation volume were established for both prismatic and basal slip modes by fitting a crystal plasticity finite element model to the lattice strain relaxation responses measured along the loading axis for three strong reflections. Prismatic slip was the easier mode having both a lower critical resolved shear stress ($\tau_c^{basal} = 252$ MPa and $\tau_c^{prism} = 154$ MPa) and activation energy ($\Delta F_{basal} = 10.5 \times 10^{-20}$ J $= 0.65$ eV and $\Delta F_{prism} = 9.0 \times 10^{-20}$ J $= 0.56$ eV). The prism slip parameters correspond to a stronger strain rate sensitivity compared to basal slip. This slip system dependence on strain rate has a significant effect on stress redistribution to hard grain orientations during cold dwell fatigue.

# 1. Introduction

Titanium alloys are lightweight high-strength materials that have been widely applied in the aerospace sector, especially in aero-engines components such as fan blades and compressor disks, often subjected to extreme mechanical loading but at relatively low temperatures [1]. In many applications, the service life of a titanium alloy component is highly dependent on its fatigue properties. An important aspect of fatigue behaviour of Ti alloys is called "cold dwell fatigue" [2]. Unlike most metals where creep is usually considered at high temperatures (when $T>0.3T_m$ where $T_m$ refers to the absolute melting temperature of the metal) [3], titanium and its alloys creep at ambient temperature. If fatigue cycles include a hold at peak loads, fatigue life of Ti alloys can be drastically reduced [4]. Whilst this phenomena has been known for a number of decades [5], the mechanisms that control this effect remain poorly explained. Understanding such failure mechanisms is extremely important for both accurate fatigue life prediction of safety critical components [6], and to inform the process design of future titanium alloys.

Prior modelling studies of dwell fatigue in titanium alloys indicate that the cold dwell effect is controlled by load shedding between "soft" and "hard" grain pairs [7,8], where the "soft" grain has its crystallographic c-axis approximately perpendicular to the loading direction while a "hard" grain has its c-axis approximately parallel to the loading direction. Characteristic of the HCP α-titanium crystal structure, plastic slip systems in the "soft" grain will be activated but within the "hard" grain no easy slip systems exist and the elastic stiffness is at its highest [9,10]. During a dwell period, localised time dependent plasticity in the soft grain leads to load shedding as dislocations accumulate and glide. Due to grain compatibility, the soft grain may impose an increased stress in a neighbouring hard grain [11,12]. With multiple loading cycles, the stress in the hard grain incrementally increases until it fractures. Fractography has revealed facet formation associated with these hard grains, where the facets themselves were typically parallel to the basal plane [5,13]. Such mechanism, however, has not been satisfactorily confirmed with experimental observations.

To make progress beyond present understanding of dwell fatigue behaviour, it is necessary to understand the different deformation behaviour in grains with different crystal orientations (i.e. 'hard' and 'soft' grain orientations) and especially the plastic deformation during the stress dwell period. To achieve this, it is important to quantify the plasticity processes as a function of time. Specifically, this must describe the dislocation processes on prismatic and basal slip systems with respect to their critical resolved shear stresses. Direct slip system measurements via observations from micro-scale single crystal mechanical testing [14] combined with crystal plasticity finite element analysis (CP-FEA) allow slip law parameters to be extracted. Gong and Wilkinson [15,16] conducted single crystal micro-cantilever bend tests to quantify the behaviour of different slip systems, keeping rate effects constant to determine effects of slip system (crystal orientation), solute and beta phase strengthening. Rate effects have received less attention, but Jun, Armstrong and Britton [17] used nanoindentation to show grain orientation dependent rate sensitivity in Ti6242, but not in the higher Mo content Ti6246 alloy. Jun *et al* [14] continued this study using in situ compression testing of micro-pillars cut within the $\alpha$ phase of Ti6242 showing the strain rate sensitivity exponent was significantly higher (m≈0.07) for prism slip than for basal (≈0.03). Subsequent, finite element simulation of these tests by Zhang *et al* [18] attempted to account for contributions to the measured indenter displacement from effects such as the punching in of the pillar into the substrate and the deformation of the sample mounting adhesive. This led to a different conclusion that the basal slip system had a higher strain rate sensitivity. Such an approach has the benefit of isolating the response of different slip systems, but is time and labour intensive and analysis has led to somewhat ambiguous conclusions.

More recently, there have been several studies on the deformation of α–titanium using synchrotron diffraction[19–25]. Such experiments enable direct assessment of lattice strains from grain families with common grain orientations from polycrystal sample. By monitoring the lattice strain response, it is possible to infer the behaviour of different slip systems.

In the current study, time dependent plasticity (stress relaxation) behaviour of commercially pure titanium (CP-Ti) was observed by synchrotron X-ray diffraction. Measured lattice strains from multiple lattice families were compared with and calibrated to simulated lattice strains from crystal plasticity finite element (CPFE) simulations using methods established in prior work [26]. The objective of this work was to quantify the key parameters controlling the time dependent plasticity of CP-Ti and develop a suitable model that enables improved quantitative prediction of cold dwell fatigue in Ti.

## 2. Materials and methods

### 2.1 Materials

The raw material used in this work was CP-Ti grade 4 rod provided by Timet with a composition indicated in Table 1.

Microstructural analysis was firstly conducted. Samples were cut from the rod then ground and polished down to a colloidal silica finish. Electron backscatter diffraction (EBSD) maps were obtained from the as received material using a Zeiss Merlin scanning electron microscope (SEM) equipped with a Bruker e-flash detector, operating at an accelerating voltage of 20 kV and a probe current of 20 nA. Figure 1a shows an example map of the microstructure, which has a mean average grain size of 17 μm. The material texture was obtained from a larger area containing over 3500 grains, shown by the pole figures in Figure 1b, where there is a preference for the c-axes to be perpendicular to the macroscopic rod axis, which is quite common in hot-rolled titanium alloys [27]. In order to ensure the easy slip system (prismatic) was activated during the stress relaxation experiment, the Electro-Thermal-Mechanical Tester (ETMT) samples were cut with the load axis along the axial direction of the rod, X0 direction in these pole figures. Due to the fact that the majority of the grains have their basal plane {0002} aligned approximately perpendicular to the tensile axis (X0 in the pole figures), the basal Schmid factor of these grains became very small, therefore prismatic slip will be the dominant slip system in this experiment. The tensile sample was 52 mm in length, and the gauge length was 16 mm with a 2 mm width and a 1 mm thickness.

Table 1. Composition of the CP-Ti grade 4 rod.

| Ti | Fe (.wt %) | O (.wt %) | N (.wt%) | Others total (.wt %) |
|---|---|---|---|---|
| Balance | 0.05-0.055 | 0.32 | 0.006 | 0.4 |

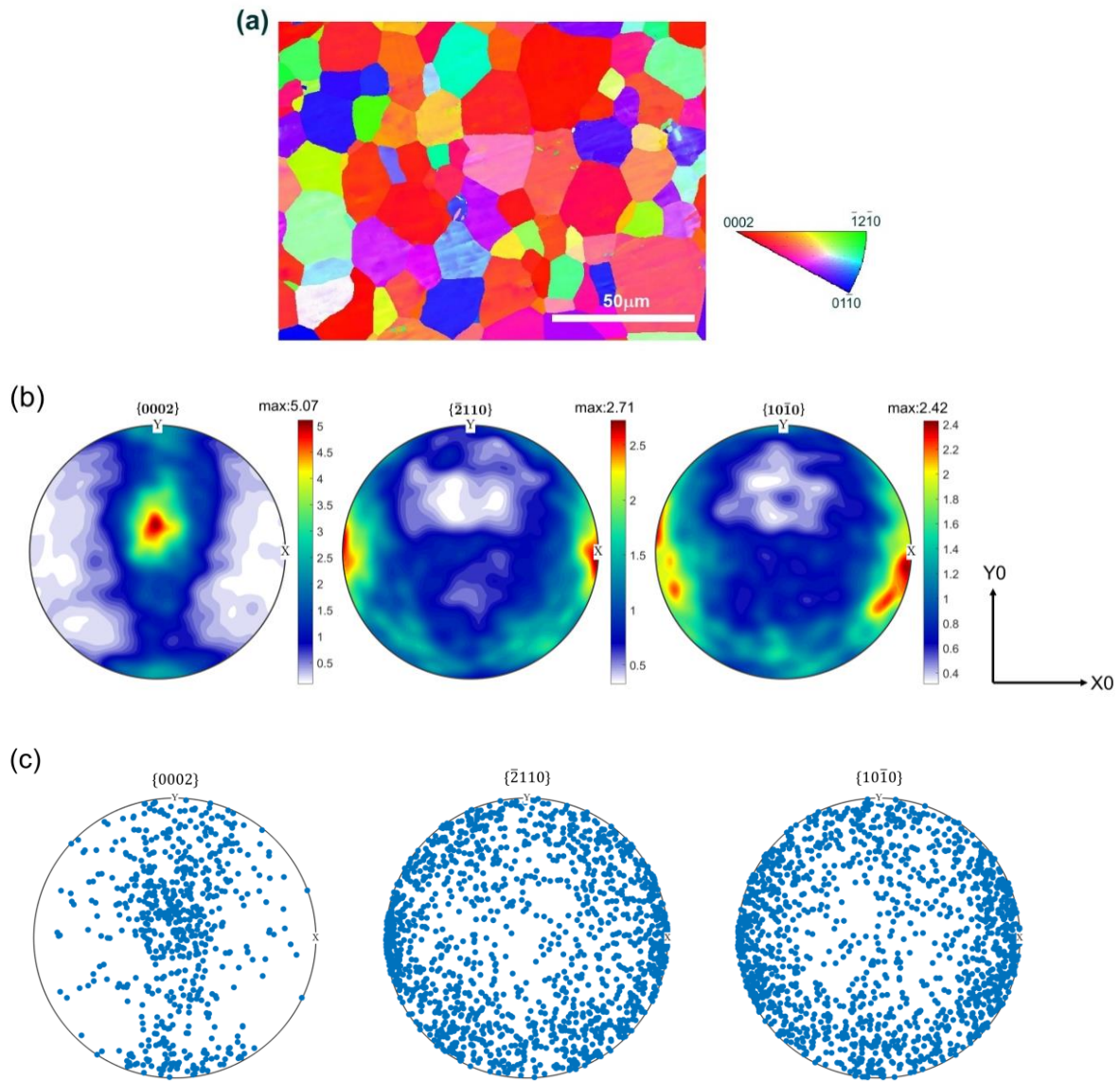

Figure 1. (a) Inverse pole figure (IPF map in radial direction of the rod) showing the grain morphology and grain size; (b) Pole figures of the raw material (X0 along the axial direction, which was the loading direction and Y0 along the tangential direction of the rod) with a 10° width used for contours plotting calculation; (c) Scatter pole figure plots of the simulated orientation used in the polycrystal finite element model.

## 2.2 Experiment

A stress relaxation experiment was carried out on beamline I12 [28] at the Diamond Light Source (experiment EE17222). An illustration of the experiment setup is shown in Figure 2.

The beamline was configured for diffraction, operating with a monochromatic beam at 79.79keV, verified with a $CeO_2$ calibration standard, and with an incident beam size of 1 x 1 $mm^2$. Dogbone shaped Ti specimens were placed in the path of the incident beam, enabling Debye Scherrer diffraction rings to be collected in transmission using a 2D Pixium area detector. The exposure time for each frame was 1 second and therefore the acquisition rate for the diffraction rings was 1 frame per second. The sample to detector distance was set and calibrated to 1097 mm. The specimens were deformed using an Instron Electro-Thermal-Mechanical Tester (ETMT) system whilst collecting Debye Scherrer diffraction data. Each diffraction pattern was recorded as a 16-bit image and synchronised with the mechanical data of the ETMT for offline analysis. Prior to each mechanical test, the sample was centred on the incident beam path using radiographic imaging to assist alignment.

The ETMT was used to subject the Ti samples to pre-determined load-hold cycles. Prior to in-situ testing, target loads were obtained from trial tests. The target loads corresponded to a stroke displacement at which the yield stress had just been exceeded. For in-situ testing, the sample was loaded at a constant load rate of 7.2 $Ns^{-1}$ until the macroscopic yield point had been exceeded by a small plastic strain. This corresponded to a strain rate of $3.38 \times 10^{-5}$ $s^{-1}$ during the initial elastic part of the loading phase. Once the target load had been reached, the sample displacement (and thus the total strain) was then held constant for a period of 5 minutes, during which the macroscopic stress was found to relax. After the 5 minutes hold, the load was incremented to a level a little below the peak load used at the start of the previous stress relaxation cycle. This load increment procedure follows that of Wang *et al* [29] and corresponds to an elastic deformation of the sample, allowing the stress relaxation at two stress levels to be compared with a negligible change in dislocation substructure. The sample was again held at a constant displacement for a further period of stress relaxation. This load-then-hold cycle was repeated for a total of 5 stress relaxation periods with diffraction patterns collected throughout (shown in Figure 3). During these relaxation periods, the stress level swept through a range of values from 450 MPa to 535 MPa, and as the plastic strains remain relatively small, changes in the mobile dislocation density $\rho$ should also be small. The elastic reloads were introduced to verify this.

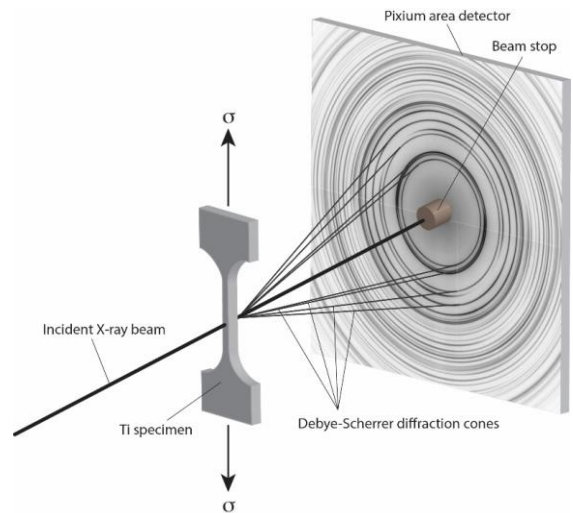

Figure 2. Schematic of the synchrotron diffraction experiment setup, the loading direction was along the axial direction of the as-received CP-Ti rod.

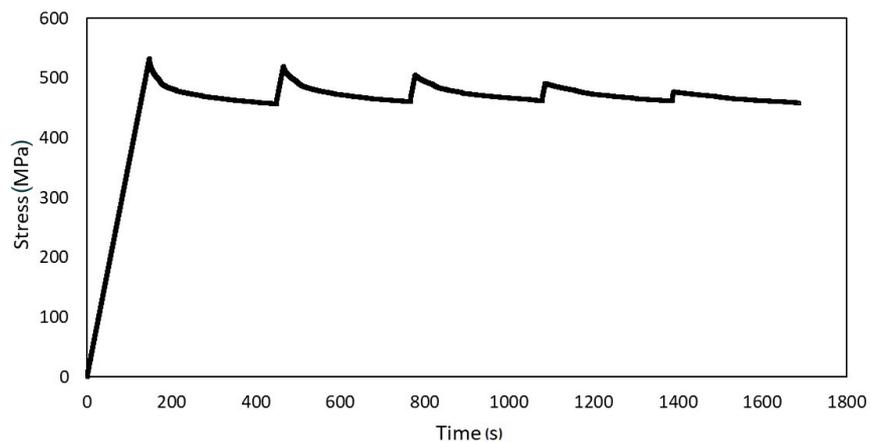

Figure 3. Loading cycles of a stress relaxation test.

The Data Analysis WorkbeNch (DAWN) [30] was used for diffraction pattern data reduction. The 2D diffraction rings were fitted with ellipses; value of the major and minor radii were determined, where the former was near parallel to the horizontal and the latter was near parallel to the vertical axis. The radius and thus the scattering vector, q, at any azimuth angle around a diffraction ring could then be found from the size and alignment of the major and minor axes. Ellipse fitting was applied to the first 21 rings with ascending ring radius, as shown in Figure 4 (with loading along the vertical axis). For further details of the data reduction process, the reader is referred Filik *et al* [31]. The scattering vectors were used to obtain the temporal lattice spacings by $d^{hkil} = 2\pi/q$. As the sample was deformed, lattice strain, $\varepsilon^{hkil}$, was subsequently calculated using $\varepsilon^{hkil} = (d^{hkil} - d_0^{hkil})/d_0^{hkil}$, where $d_0^{hkil}$ is the initial lattice spacing for plane *hkil*.

To find a consistent set of stress-free lattice parameters, *a* and *c,* the following procedure was used. For each diffraction ring, a temporal $d_0^{hkil}$ value was obtained by performing a linear fit to the *d* spacing vs. time plot (as shown in Figure 5a) during the initial elastic loading, then extrapolating the fitted linear function to time zero. These temporal $d_0^{hkil}$ values from different diffraction planes were then fitted to the crystal geometry equation for HCP:

$$\frac{1}{d^{hkil}} = \frac{3(h^2+hk+k^2)}{4a^2} + \frac{l^2}{c^2} \tag{1}$$

where *a* and *c* are the lattice parameters of HCP crystal, *h, k* and *l* are the Miller Indices of a crystal plane. Therefore, the non-deformed lattice parameters of this CP-Ti sample were found to be $a = 2.9512 \pm 0.002$ Å and $c = 4.6861 \pm 0.004$ Å (see Figure 5b). The initial stress-free lattice spacing, $d_0^{hkil}$, for each plane family was then obtained using these values of *a* and *c* and the formula above.

The diffraction line profiles shape for some reflections was investigated, however no measureable changes in peak width during the tests were found. This was due to a significantly larger contribution of broadening from the instrument than any broadening from the sample; making determination of the latter prohibitive.

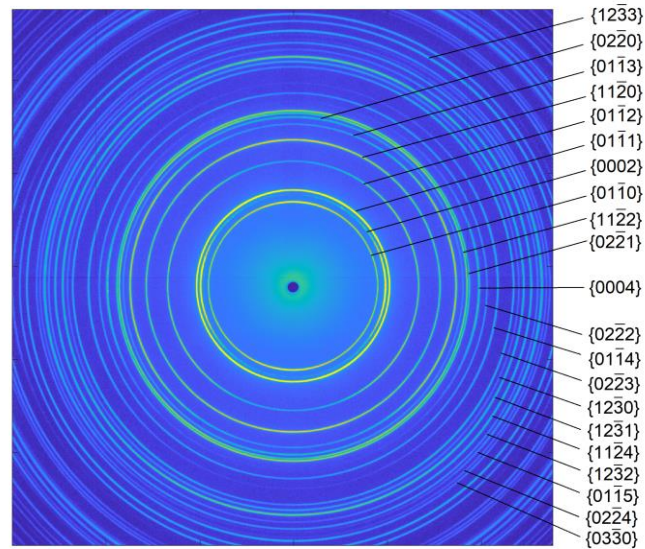

Figure 4. Debye Scherrer diffraction patterns of CP-Ti at room temperature (the vertical direction of the diffraction patterns was the axial direction and the horizontal direction is the transverse direction of the CP-Ti rod).

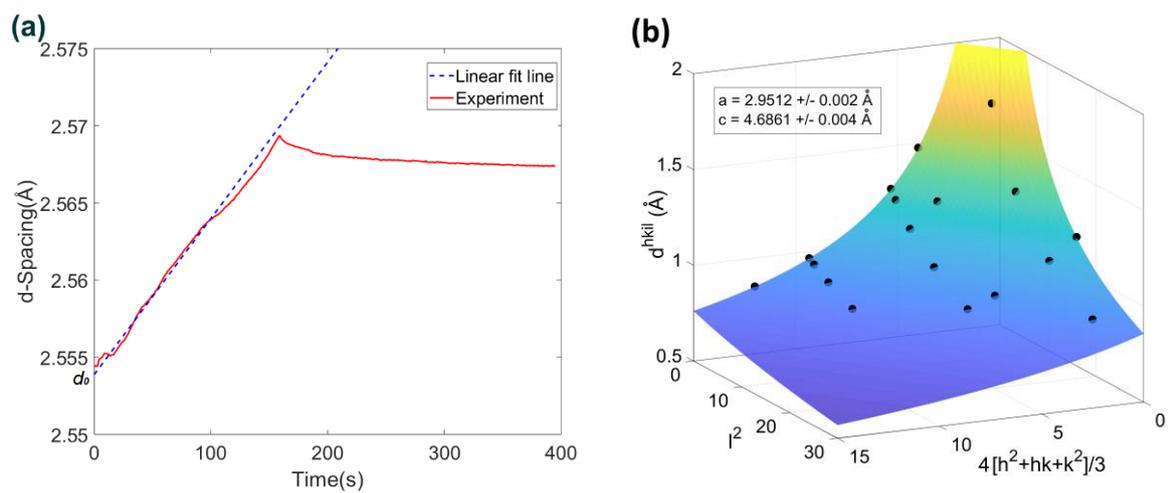

Figure 5. (a) Linear fit of $d$ spacings during elastic loading to determine temporal $d_0^{hkil}$ for each plane family (the $\{01\bar{1}0\}$ plane is shown here as example); (b) Non-linear regression fit of the crystal geometry equation for a HCP crystal.

## 2.3 CPFE model

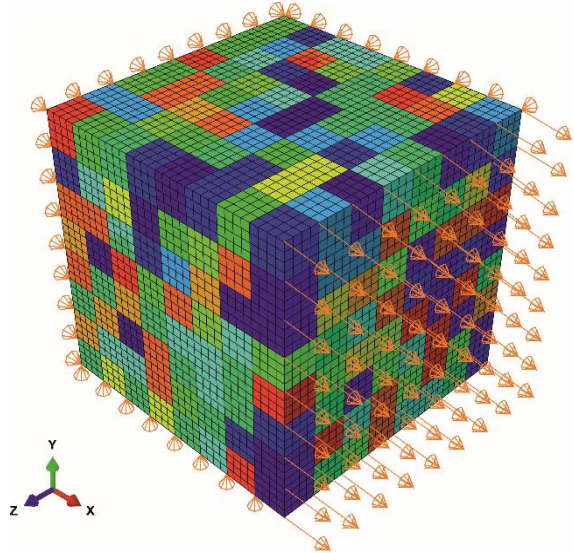

Figure 6. Finite element mesh of the RVE used with the CPFE model. The colours represent the grain orientation shown in figure 1, and arrows indicate the uniaxial boundary conditions.

The CP-Ti material was modelled as a single phase, α, with a HCP crystal structure. The elastic parameters used are indicated in Table 2. Dislocation slip occurs predominantly with <a> type Burgers vector on the basal and prismatic slip planes in Ti alloys [25–27], while the CRSS required to activate <c+a> 1st & 2nd order pyramidal slip can be about three time the value of basal slip [7,10]. Since in this experiment, the sample was loaded just beyond the yield point, the <c+a> slip was relatively difficult to activate, hence, little activity of pyramidal slip was expected.

Table 2. Parameters for polycrystalline Ti model under uniaxial loading at room temperature[28,29].

| Moduli (GPa) and Poisson's ratio | | | | | |
| --- | --- | --- | --- | --- | --- |
| $E_{11}$ | $E_{33}$ | $G_{12}$ | $G_{13}$ | $v_{12}$ | $v_{13}$ |
| 84.7 | 119.8 | 29.0 | 40.0 | 0.46 | 0.22 |

A crystal plasticity model was implemented using the commercial finite element (FE) software Abaqus 2016 through a user material (UMAT) subroutine developed by Tarleton *et al* [37] based on a formulation initially proposed by Dunne *et al* [35]. Briefly, the kinematics are governed by the deformation gradient (**F**) which is decomposed multiplicatively into an elastic (**F**$^e$) and plastic (**F**$^p$) part. The rate of change of the plastic deformation gradient ($\dot{\boldsymbol{F}}^p$) required to calculate the increment in plastic strain over a small time increment can be written in terms of the plastic velocity gradient (**L**$^p$) which comes from the linear sum of shear strain rates over the possible crystallographic slip systems.

The material properties were next introduced through the constitutive law relating the shear strain rate on each slip system to the local stress state. We use the constitutive law that was physically derived by Dunne *et al* [35], in which the slip rate, $\dot{\gamma}^\kappa$, on a slip system $\kappa$ is given by:

$$\dot{\gamma}^\kappa = \rho b^2 v \exp\left(-\frac{\Delta F^\kappa}{k_B T}\right) \sinh\left(\frac{(|\tau^\kappa| - \tau_c^\kappa)\Delta V^\kappa}{k_B T}\right) \operatorname{sgn}(\tau^\kappa) \qquad (2)$$

where $\rho$ is the density of gliding dislocations (here taken to be a constant 5 μm$^{-2}$ following [1,28,31]), $b$ is the magnitude of the Burgers vector (from the measured *a* and *c* values), $v$ is the jump frequency (i.e. attempts of the dislocations to exceed their energy barriers was taken to be $10^{11}$ Hz, following [1,28,31]), $k_B$ is the Boltzman constant and $T$ is the absolute temperature ($T = 298$ K throughout). The remaining three terms are material specific parameters that govern dislocation motion and are the variables to be obtained from this study: $\Delta F^\kappa$ is the slip system dependent activation energy, which corresponds to the energy barrier that a dislocation on a slip system $\kappa$ needs to overcome to continue gliding beyond the pinning obstacle, $\Delta V^\kappa$ is the activation volume, which refers to the volume that a dislocation is swept when it passes obstacles [35] ($\Delta V^\kappa$ is typically in the range 1-100 $b^3$), and $\tau_c^\kappa$ is the critical resolved shear stress (CRSS) for slip system $\kappa$. If the resolved stress falls below CRSS, slip does not occur. In this condition $\dot{\gamma}^\kappa = 0$. The slip system dependent parameters $\Delta F$, $\Delta V$ and $\tau_c$ are fitted in this study.

A polycrystal finite mesh containing 512 grains with 64 quadratic elements (C3D20R) per grain with 8 integration points per element was generated in Abaqus as shown in Figure 6. For simplicity the model adopted a simplified morphology approximating the grains as a regular cubic array [19,32]. The discrete orientations, as measured by EBSD, were selected at random to assign orientations grain by grain, therefore the overall texture in the model was approximately the same as the real sample (as shown in Figure 1c). The boundary conditions are illustrated in Figure 6, a uniaxial displacement of 1.4 µm was applied to the right surface over 150s (the total length of one edge of the model was 240 µm, therefore the displacement corresponded to a strain rate of $3.8 \times 10^{-5}$ s$^{-1}$), and then the displacement was held fixed along the load axis for 300 s while the left, bottom and back surfaces were constrained in the X, Y and Z directions respectively.

The simulated elastic strains, were compared with elastic lattice strains obtained from diffraction measurements. The Green-Lagrange elastic strain tensor in the sample frame, given by $\boldsymbol{E}_e$, is rotated into the lattice orientation frame through the rotation matrix $\boldsymbol{R}$ to give the lattice strain $\boldsymbol{E}^0$ as follows [39]:

$$\boldsymbol{E}^0 = \boldsymbol{R}^T \boldsymbol{E}_e \boldsymbol{R} = \frac{1}{2} \boldsymbol{R}^T (\boldsymbol{F}_e^T \boldsymbol{F}_e - \boldsymbol{I}) \boldsymbol{R} \qquad (3)$$

As illustrated in Figure 7, diffraction from the $\{01\bar{1}0\}$ planes, for example, with diffraction observed along the vertical axis and transverse axis have diffraction vectors, $q$, approximately parallel (with 10 degree tolerance) to the $E^0_{yy}$ and $E^0_{xx}$ components respectively. For grains that satisfy this condition, the $E^0_{yy}$ and $E^0_{xx}$ components of the $\boldsymbol{E}^0$ matrix were averaged from all the elements within these grains and directly compared with the experimental lattice strains.

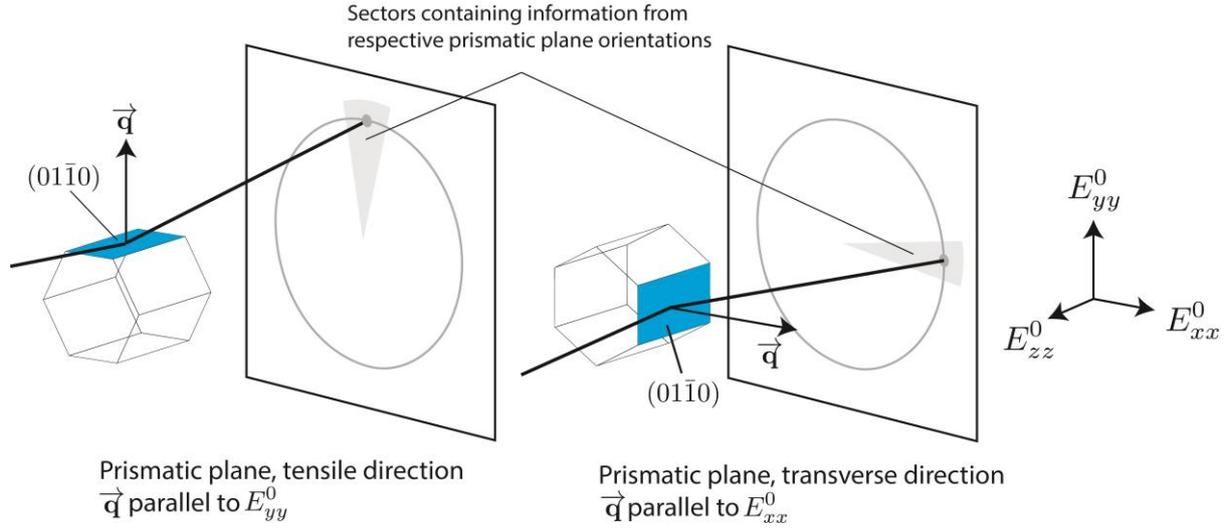

Figure 7. Assessed prismatic plane diffraction orientations.

The geometrically necessary dislocation (GND) density was calculated by solving the Nye dislocation tensor, $\boldsymbol{\alpha}^{Nye}$ [40]:

$$\boldsymbol{\alpha}^{Nye} = \sum_\lambda (\boldsymbol{b}^\lambda \otimes \boldsymbol{\rho}^\lambda) \qquad (4)$$

where $\boldsymbol{b}$ is the Burgers vector, $\boldsymbol{\rho}$ is the GND density and $\lambda$ is a general slip system. This tensor can be found from the curl of the plastic deformation gradient, $\boldsymbol{F}^p$, using the following convention:

$$\alpha^{Nye}_{km} = \epsilon_{ijm} F^p_{kj,i} \qquad (5)$$

A linear operator $\boldsymbol{A}$ is introduced (9×$j$ matrix, for $j$ types of dislocations), where the $j^{th}$ column contains the dyadic product of the Burgers vector and line direction of the $j^{th}$ dislocation type. Therefore, the $j$ types of dislocation densities are represented in a column vector, $\boldsymbol{\rho}$.

$$\boldsymbol{A\rho} = \boldsymbol{\alpha} \qquad (6)$$

Since generally $j > 9$ there is no unique solution for $\rho$. Instead, with knowledge of $\boldsymbol{\alpha}$ and $\boldsymbol{A}$, it is straightforward to find the right pseudo inverse of $\boldsymbol{A}$ to give an estimate $\boldsymbol{\rho}$

$$\boldsymbol{\rho} = \boldsymbol{A}^T(\boldsymbol{A}\boldsymbol{A}^T)^{-1}\boldsymbol{\alpha} \qquad (7)$$

For further details, the reader is referred Das, Hofmann and Tarleton[37].

## 3. Results & Discussion

3.1 Macroscopic stress relaxation

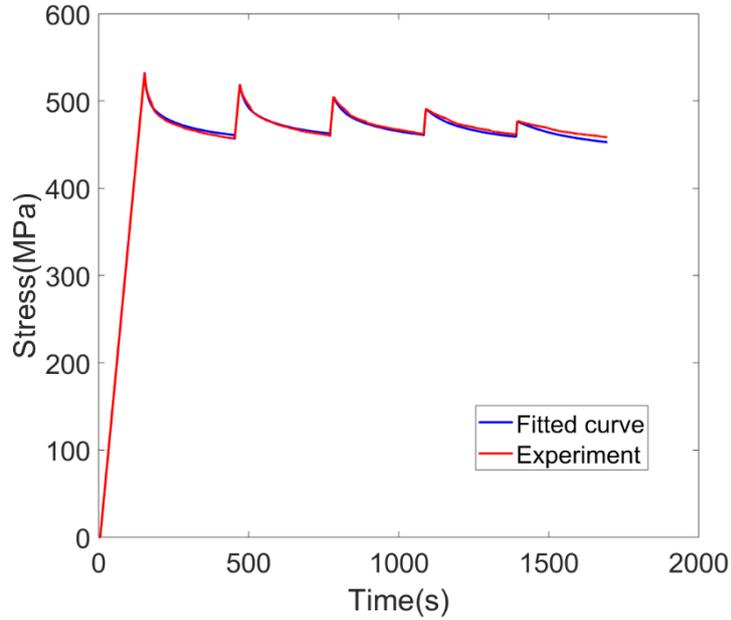

Figure 8. Macroscopic stress relaxation curve fitting.

The macroscopic stress relaxation response, calculated from the measured force from the ETMT load cell is shown in Figure 8. It was firstly fitted using the same form of constitutive law as Equation 2, but instead linking the macroscopic plastic normal strain rate to the macroscopic normal stress:

$$\dot{\varepsilon} = \rho b^2 v \exp\left(-\frac{\Delta F}{kT}\right) \sinh\left(\frac{(\sigma-\sigma_c)\Delta V}{kT}\right) \quad (8)$$

$$\sigma_i \rightarrow \Delta\varepsilon = \dot{\varepsilon}\Delta t \rightarrow \Delta\sigma = E\Delta\varepsilon \rightarrow \sigma_{i+1} = \sigma_i + \Delta\sigma$$

where $\dot{\varepsilon}$ is the macroscopic strain rate, $\sigma$ is the applied stress and $\sigma_c$ is the critical stress. With the initial stress obtained from the experiment and Equation 8, the macroscopic strain increment, $\Delta\varepsilon$, was calculated for a small time increment $\Delta t = 0.2s$. $\Delta\varepsilon$ was then multiplied by the average Young's modulus for CP-Ti grade 4 ($E = 105$ GPa) [41] to obtain an estimate of the average stress increment, $\Delta\sigma$, which was then used to update the stress for the next time increment. This process was repeated in order to reconstruct a stress relaxation curve over the displacement hold period. With the aid of fitting tools in Matlab, the best-fit relaxation curve (blue line in Figure 8) was found and the optimal fitting parameters $\Delta F^{macro}, \Delta V^{macro}$ and $\sigma_c^{macro}$ obtained are listed in Table 3. These values were then used as the initial values in the crystal plasticity simulations.

Table 3. Fitted parameters obtained from the best-modelled relaxation curves.

|  | Parameters |
| --- | --- |
| $\sigma_c^{macro}$ | 449.8 MPa |
| $\Delta F^{macro}$ | $10.17 \times 10^{-20}$ J |
| $\Delta F^{macro}$ | 0.63 eV |
| $\Delta V^{macro}$ | 9.93 $b^3$ |
| Fitting error | 0.61% |

This numerically fitted parameters are broadly similar to those used by other workers in CPFE modelling, where Dunne et al [35] stated that $\Delta F = 0.73$ eV for a Ti-6Al alloy, Zheng et al [1] used $\Delta F = 0.6$ eV and $\Delta F = 0.66$ eV for Ti-6242 and Ti-6246 alloys respectively while $\Delta V = 0.5b^3$ for both alloys and Zhang et al [38] demonstrated that $\Delta V$ is different for $\alpha$ and $\beta$ phases ($\Delta V_\alpha = 11.97b^3$ and $\Delta V_\beta = 0.0021b^3$) in a Ti-6242 alloy, while all of these studies used the same prefactors for Ti alloys as this work ($\rho = 5$ μm$^{-2}$, $b = 0.295$ nm and $\upsilon = 10^{11}$ Hz). This indicates that this numerically fitting macroscopic stress relaxation curve method is able to give a corroborating prediction of these key parameters of the alloy tested. Therefore, these values stated in Table 3 were then used as the initial values in the crystal plasticity simulations. The ability to find a good fit that is consistent across the five repeated periods of stress relaxation indicates that strains are sufficiently low that the evolution of dislocation microstructure through the test can be neglected.

## 3.2 Development of lattice strain

Deformation of α-Ti is highly anisotropic both elastically and plastically, resulting in deformation properties that vary significantly depending on the crystallographic orientation of a grain. In order to quantify the crystallographic orientation, the concept of a c-axis declination angle is introduced (as shown in Figure 9), which is the angle, $\varphi$, between the load axis and the c-axis of the α-Ti HCP crystal. When a grain with prismatic $\{01\bar{1}0\}$ planes satisfy the diffraction condition with its diffraction vector, $q$, approximately parallel to the load axis, the c-axis declination angle is 90°. Grains in this orientation are referred to as 'soft'. When a grain with its basal $\{0002\}$ planes satisfy the diffraction condition, again with the diffraction vector, $q$, approximately parallel to the load axis. Thus, its c-axis declination angle is 0° and the grain is considered to be 'hard'[7]. Other than these two special crystal orientations, the c-axis declination angle can be calculated from the angle between the c-axis and the normal of the plane satisfying the diffraction condition along the loading direction. As the c-axis declination angle varies between 0° to 90°, the grain orientation changes gradually from 'hard' to 'soft'.

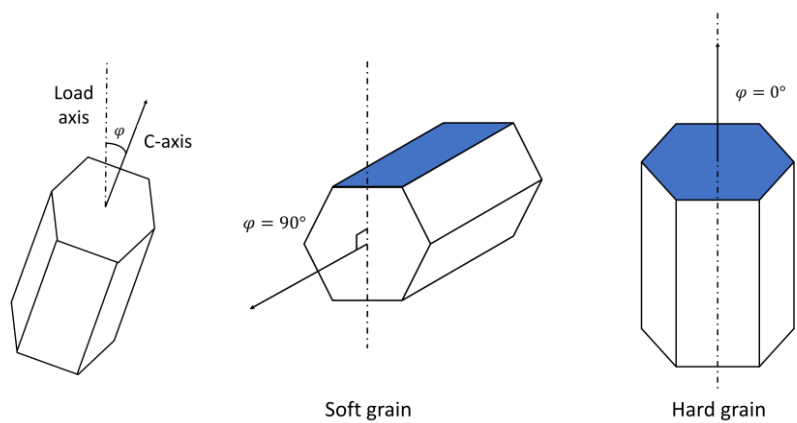

Figure 9. Schematic of HCP crystals that illustrates the c-axis declination angle, $\varphi$. Example grains in 'soft' and 'hard' orientations are shown.

The development of axial lattice strains is shown in Figure 10a and 10b for $\{01\bar{1}0\}$ and $\{12\bar{3}0\}$ planes as examples, measured from the vertical axis (loading direction) of the diffraction patterns. Linear fitting was applied to the initial stage (elastic part) of the macroscopic stress vs. lattice strain plots. The gradient of the fitted function was used to measure the stiffness of each corresponding orientation. The different plane family stiffnesses against their c-axis declination angle are shown in Figure 10c. It is shown that as the declination angle increases, the stiffness decreases (from 135 GPa to 96 GPa). This neglected the two data points at $\varphi = 0°$, which were deemed unreliable as they were obtained from data with high noise. This trend agrees with the elastic modulus variation with declination angle for titanium, where the elastic modulus along the c-axis of the $\alpha$-titanium unit cell is about 1.4 times higher the value of that along the a-axis (see Table 2)[28,29,34]. Applying the same method to the transverse lattice strain, the transverse stiffness was calculated (Figure 10d). Showing a reversed behaviour to the axial stiffness, the transverse stiffness increases with the increasing declination angle. By combining the axial and transverse stiffness, the Poisson's ratio was determined (Figure 10e), which increases from 0.2 to 0.46 as the declination angle increases from 0° to 90°, this agrees well with the literature values [29,35].

By extending the linear fitted lines backward in Figure 10a and 10b, the intersections of these lines with x and y axes were calculated to estimate the residual lattice strain and stress. As shown in Figure 10f and 10g, the residual strain and stress are also anisotropic, due to the combined action of thermal and mechanical treatments during processing [9,10,36].

Gradient changes were observed during the loading process (Figure 10a and 10b), which appear at $\approx 200$ MPa for the $\{01\bar{1}0\}$ lattice planes and $\approx 155$ MPa for the $\{12\bar{3}0\}$ lattice planes. Different lattice planes start to deviate from a linear macroscopic stress-lattice strain variation, indicating plastic deformation. For most of the plane families, grains have a lower increment in lattice strain with respect to macroscopic stress (loading curves deviate upwards, see Figure 10a), as a result some plane families have the opposite response (Figure 10b) so that the whole system is in static equilibrium. The initiation of plastic deformation is related to the activation of slip systems. At the stress level for this experiment, the prismatic and basal slip systems are more important than the hard pyramidal slip system (recall section 2.3 and Appendix C), therefore the maximum Schmid factor among prism and basal slip systems were calculated for the 21 plane families and correlated to this onset of plasticity stress. Figure 10h shows the onset of plasticity stresses of 16 plane families whose gradient change can clearly be observed against their maximum Schmid factor for prism and basal slip. It is found that grains with a higher maximum Schmid factor tends to activate plastic deformation at a lower macroscopic stress.

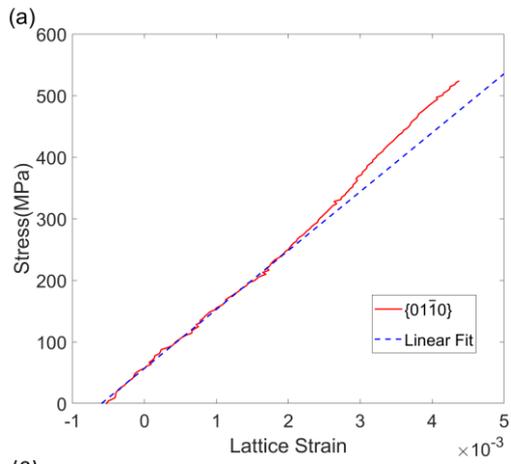
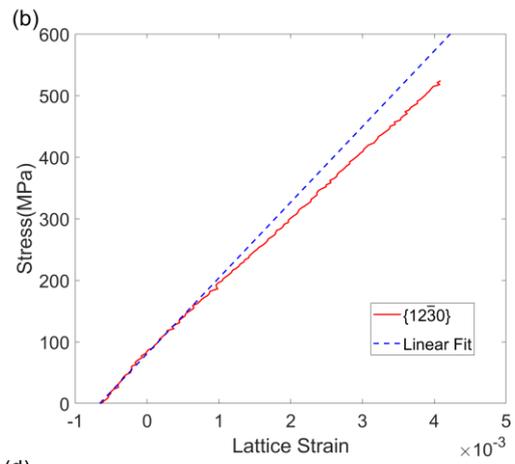
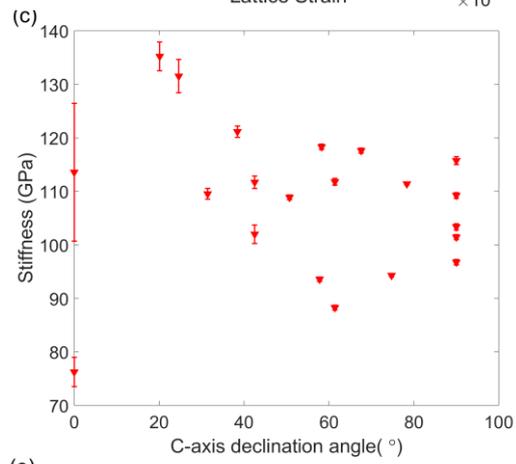
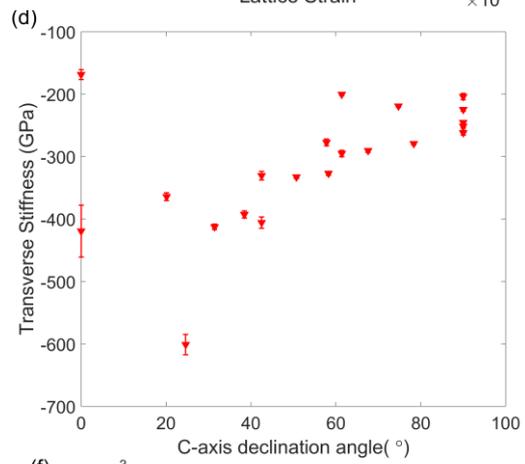
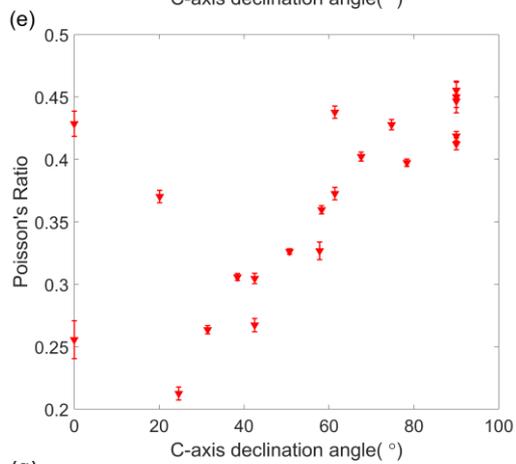
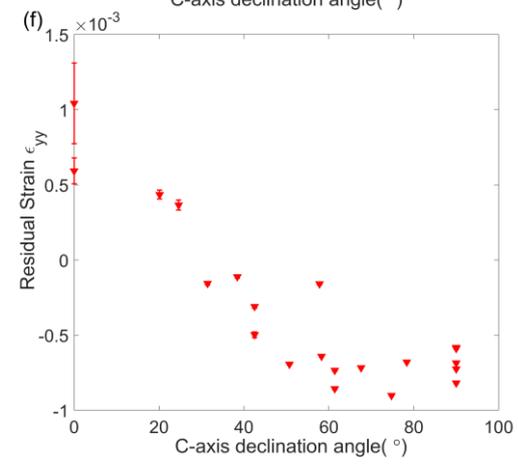
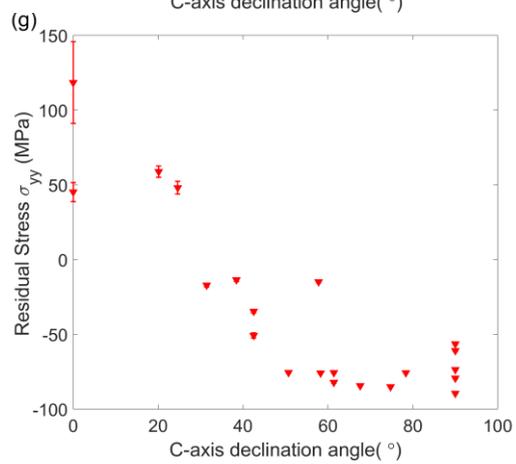
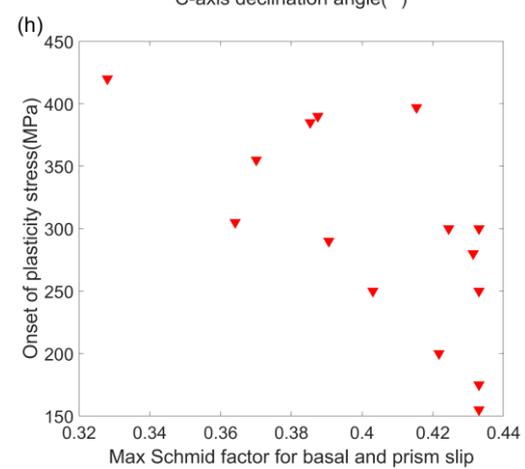

Figure 10. (a) Development of the elastic lattice strain in the loading direction for the $\{01\bar{1}0\}$ plane; (b) Development of the elastic lattice strain in loading direction of the $\{12\bar{3}0\}$ plane; (c) Axial stiffness vs c-axis declination angle; (d) Transverse stiffness vs. c-axis declination angle; (e) Poisson's ratio vs. c-axis declination angle; (f) Axial residual strain vs. c-axis declination angle; (g) Axial residual stress vs. c-axis declination angle; (h) Onset of plasticity stress vs. maximum Schmid factor for basal and prism slip.

3.3 Modelling results

The lattice strains developed during the first 5-minute whilst the sample displacement was held were used to calibrate the model. Simulated lattice strains were extracted to fit with three important reflections; these comprised the prismatic plane $\{01\bar{1}0\}$, the first order pyramidal plane $\{01\bar{1}1\}$ and the second order pyramidal plane $\{11\bar{2}2\}$. The parameters related to dislocation motion in the slip law (Equation 2) were adjusted until the simulated axial lattice strains fitted well with the experimental axial lattice strain of these three reflections. The basal plane $\{0002\}$ was not available in the simulations due to the texture of the sample, giving few grains diffracting along the loading axis or the axial direction of the CP-Ti rod (recall Figure 1b and 1c). Residual strain was not included in the simulations; details of justification of this assumption can be found in Appendix A. It was assumed that little <c+a> pyramidal slip will be activated in this small strain deformation, therefore parameters $\Delta V$ and $\Delta F$ for pyramidal slip were fixed to the parameters calculated from the macroscopic stress relaxation curve in section 3.1. The CRSS of pyramidal slip was fixed to be 3 times the value of CRSS of basal slip [7,10] (justification of these assumption can be found in Appendix C). Hence, there are six parameters ($\Delta V$, $\Delta F$ and $\tau_c$ for prismatic and basal slip systems) to calibrate in order to obtain an optimal fit.

Parametric studies of the effects of fitting parameters on the lattice strain relaxation curves generated by the simulation showed that the outputs are not so sensitive to $\Delta V$ changes compared to the other fitting parameters (details can be found in Appendix C). This applies for the relatively low strain rates seen in our tests, but of course may not hold at higher strain rates where $\Delta V$ might be expected to have a more significant impact on behaviour. Therefore, the value of $\Delta V$ was fixed to the value obtained from macroscopic stress relaxation for both basal and prismatic slip systems. The lattice strain of the $\{01\bar{1}0\}$ planes is thought to be dominated by prismatic slip as the basal plane for the grain family is nearly parallel to the loading direction, and so the basal slip Schmid factor is approximately zero. Fitting the model output to the $\{01\bar{1}0\}$ lattice strain variation first allows an initial focus on only two unknown parameters i.e. $\Delta F$ and $\tau_c$ for prism slip. Changing $\Delta F$ will result in a clear shape change of the strain relaxation curve (i.e. the amount of relaxation within the hold period) while $\tau_c$ mainly affects the lattice strain level (an increase in $\tau_c$ shifts the whole relaxation curve upwards and vice versa). The $\{01\bar{1}0\}$ lattice strain result was fitted first to confirm the parameters for prismatic slip. This was followed by consideration of the $\{01\bar{1}1\}$ and $\{11\bar{2}2\}$ lattice strains, which were affected by parameters describing both basal and prismatic slip. Finally, the simulated macroscopic stress relaxation curve was also compared with the experiment curve. Figure 11a shows the best-fit lattice strain relaxation curves along the loading direction for the $\{01\bar{1}0\}$, $\{01\bar{1}1\}$ and $\{11\bar{2}2\}$ planes. The transverse lattice strains (not used in the fitting) are compared between the experiment and simulation in Figure 11b. The macroscopic stress relaxation also shows a good fit with the experimental results (Figure 11c). It is noted that whilst the prefactor $\rho$ (mobile dislocation density) was fixed through all five relaxation cycles, the simulation results show good agreement with the experimental results. This verified that the change in mobile dislocation density remains relatively small (recall section 2.2).

The corresponding slip law parameters for these best fits are summarised in Table 4. The CRSS ratio between basal and prismatic slip is a particularly important factor for the deformation of HCP metals. For both basal and prismatic slip to be activated during deformation, a ratio should be close to 1.0[45], which is also beneficial for ductility and to reduce the effect of texture. The ratios have been well studied and values from the literature are listed in Table 5 in comparison to our result.

The values of both $\tau_c^{basal}$ and $\tau_c^{prism}$ determined in this work are higher than the values in table 5 (except the one tested on a single crystal). This is because materials used in the present work is CP-Ti grade 4, which has a much higher oxygen content compared to others. Oxygen, as an interstitial atom, and has very pronounced strengthening effect in Ti alloys even at low concentration[38–40]. Besides this, researchers have also reported that the strengthening effect of oxygen is more pronounced on the softer prism slip systems than basal slip systems [41,42]. This explains why $\tau_c^{basal}/\tau_c^{prism}$ in this work is lower than examples reported in the literature for lower oxygen concentrations. Unlike prior modelling studies, the values obtained for the activation energy $\Delta F$ was found to be different for prismatic and basal slip systems, where $\Delta F^{basal}$ is about 17% larger than $\Delta F^{prism}$. This indicates that basal slip activation is more difficult than prismatic slip due to two factors: basal slip has a higher room temperature critical resolved shear stress and a higher energy barrier to overcome.

Table 4. Key parameters associated with dislocation motion for prismatic and basal slip

|  | $\tau_c$ (MPa) | $\Delta F$ ($\times 10^{-20}$ J) | $\Delta F$ (eV) | $\Delta V$ ($b^3$) |
| --- | --- | --- | --- | --- |
| Prismatic slip | 154 | 9.0 | 0.56 | 9.93 |
| Basal slip | 252 | 10.5 | 0.65 | 9.93 |
| Pyramidal slip | 756 | 10.17 | 0.63 | 9.93 |

Table 5. CRSS for basal and prismatic slips values for commercially pure Ti. The nominal oxygen content and average grain sizes from each study is listed.

| $\tau_c^{prism}$ (MPa) | $\tau_c^{basal}$ (MPa) | $\tau_c^{basal}/\tau_c^{prism}$ | O (wt%) | Grain size (μm) | Ref. |
|---|---|---|---|---|---|
| 120 | 182 | 1.5 | 0.16 | 40 | [51] |
| 30 | 150 | 5.0 | <0.01 | 30 | [52] |
| 60 | 120 | 2.0 | 0.17 | 80 | [45,46] |
| 68 | 175 | 2.6 | 0.12 | 50 | [55] |
| 90 | 180 | 2.0 | 0.11 | 9 | [56] |
| 181 | 209 | 1.2 | 0.07 | Single crystal | [15] |
| 96±18 | 127±33 | 1.7-2.1 | 0.17 | 100 | [45] |
| 154 | 252 | 1.63 | 0.32 | 17 | This work |

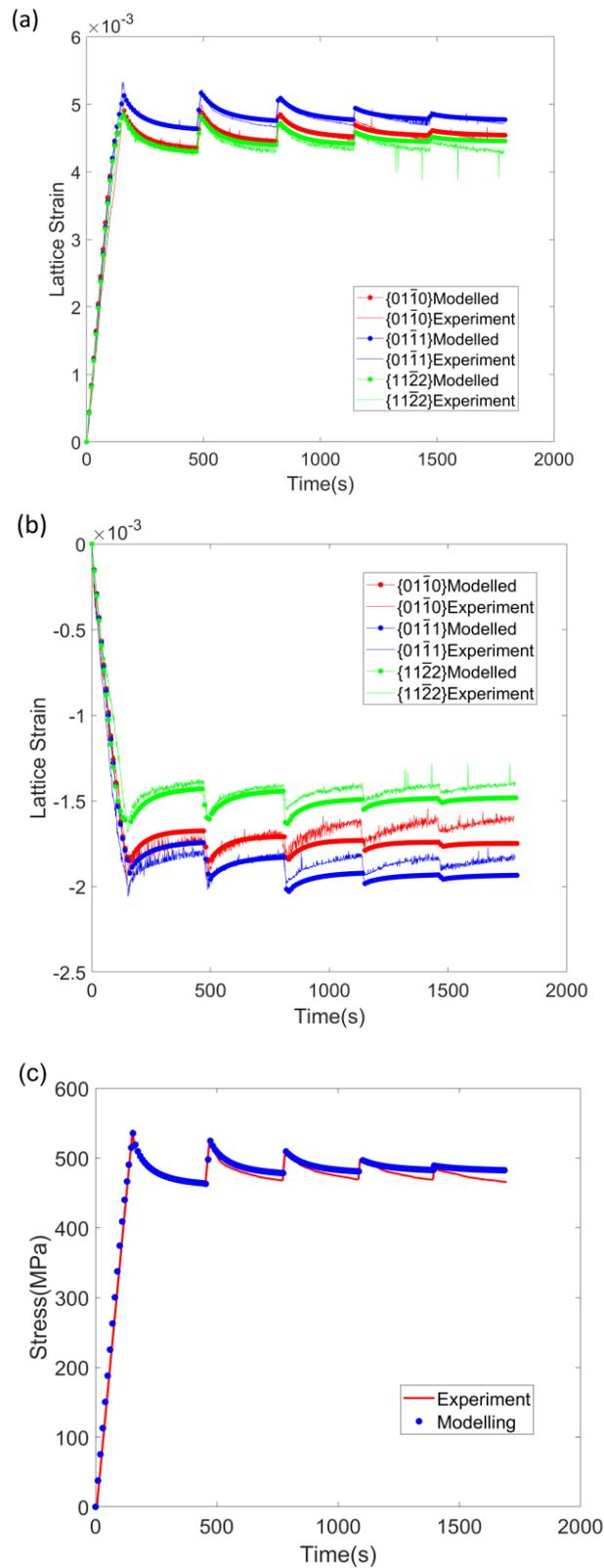

Figure 11. (a) Lattice strain evolution along the loading direction; (b) lattice strain evolution along the transverse direction and (c) macroscopic stress relaxation curves.

## 3.4 General Discussion

This study was motivated by a need to understand the technologically important cold dwell fatigue, which effects some but not all Ti alloys and is thought to be caused by time dependent plastic deformation on the softer slip systems. Significant stress relaxation at room temperature (≈0.15 of the melting point) was clearly demonstrated in the commercially pure Ti studied here and captured in lattice strain drops measured via synchrotron XRD during macroscopic strain hold periods. The extent of relaxation varies between different grain families as indicated by the increment of plastic strain inferred from lattice strain variations over the first 5 min hold period shown in Figure 12d. Deconvolving this grain orientation effect on the relaxation behaviour requires CP-FEA simulations which shows that the activation energy is higher for basal compared to prism slip. This is in addition to the expected higher CRSS for basal compared to prism slip. These combine to give a greater strain rate sensitivity for the softer prism slip system.

Results in the literature are often expressed in terms of a strain rate sensitivity parameter, $m$, which corresponds to the gradient of a plot of log ($\sigma$) against log ($\dot{\varepsilon}$) and would imply the strain rate exponent in a simple power law expression for stress. We have attempted to assess the strain rate sensitivity of different grain families by analysing the lattice strain relaxation data for individual reflections (e.g. Figure 12a). The noise in the experimental lattice strain results, increases the difficulty in calculating strain rate. The lattice strain relaxation curves were smoothened using the same method as fitting the macroscopic stress data relaxation (discussed in section 3.1). The gradient of the smoothed and fitted lattice strain relaxation curve at each second (out of total 300s) were calculated to represent the lattice strain rate at each second. Lattice strain rates, which are negative, are assumed to be matched by positive plastic creep strain rates so that the overall strain in that grain family is held constant. The stresses for that grain family are obtained from the product of the lattice strain value and the corresponding stiffness (Figure 10c). This allows creep rates and corresponding stresses to be assembled in a log-log plot (Figure 12b), the gradient of which gives the effective SRS parameter, $m$, for that grain family. The variation of grain family effective SRS values with c-axis declination angle is shown in Figure 12c, and implies that softer grains have higher effective SRS than those of harder grains. However, assuming that the external boundary condition of a fixed total strain can be imposed on individual grain families is questionable, as for the softer grains, larger plastic strains may be achieved if accommodated by smaller plastic strains in a harder neighbouring grain environment. The plastic strain rate for the softer grain families is thus likely to be underestimated in this analysis, and conversely overestimated for the harder grains. This coupling to other grains in the neighbourhood prevents sound knowledge of the boundary conditions and is what drives the need to use a CP-FEM simulation approach in which spatial variation in stress and strain evolution is accounted for.

An increment of plastic creep strain during the first macroscopic strain hold period was obtained by subtracting the lattice strain at the end of the relaxation period from the lattice strain at the beginning of the relaxation period. As shown in Figure 12d, the increment of plastic creep strain was found to be larger in softer grains than that of harder grains. The same trend was found in the plot of plastic strain increments for 21 reflections against their averaged Schmid factor for prism and basal slips, as shown in Figure 12e. Combining with the variation of SRS vs. c-axis declination angle, the relationship of the SRS and the accumulation of plastic strain during creep can be correlated. A higher SRS can be seen to result in greater plastic strain increments during cold creep of Ti.

Figure 12f shows plots of stress vs. strain rate variation given by the slip laws for prism and basal slip. We have adjusted from shear stress to normal stress by dividing by a Schmid factor of 0.5, and from shear strain rate to normal strain rate by multiplying by the Schmid factor (again taken to be 0.5). Also shown on this plot is the sample averaged response measured by the ETMT during the first 4 periods of stress relaxation, which fall between those for the basal and prism slip laws. This plot is important in showing that strain rate sensitivity can be expected to vary significantly with the strain rate range over which it is determined. In the relatively low strain rate range directly covered by the experimental data ($10^{-7}$-$10^{-5}$) the slip laws can be approximated by SRS values of 0.052 for prism, and 0.040 for basal slip. For comparison with the work by Jun *et al* [14] who reported SRS values of for 0.03 basal, and 0.07 prism slip for micro-pillars cut from single $\alpha$ phase crystals in Ti6242 measured over the strain rate domain $10^{-4}$ to $10^{-2}$. Extrapolating to this higher range of strain rates for the slip laws determined here for grade 4 CP-Ti gives SRS values of 0.039 for basal, and 0.064 for prism slip. These values for CP-Ti and Ti6242 are in remarkable agreement. It is intriguing that such similar results are found across the relatively simple and soft commercially pure Ti, and the much harder alloy with significant substitutional alloying but lower interstitial content.

The analysis process suggested very similar activation volumes for the two slip systems so that in the very high stress, very high strain rate regime the SRS of prism and basal slip should tend to the same value. The activation volume is too small (~$10b^3$) to suggest pinning by dislocation-dislocation interactions (at the assumed dislocation density of $5 \times 10^{12}m^{-2}$ the spacing would suggest an activation volume ~$1500b^3$), but significantly larger than the ~$1b^3$ expected for an atomistic diffusion process. However, in the BCC metals activation volumes of the size seen here have been ascribed to kink pair nucleation mechanisms which limit the motion of the slower screw dislocations [57]. A similar physical basis seems reasonable here. These differences in strain rate sensitivity will obviously impact on the details of how load is transferred from soft grains onto hard grains as whether slip is on a prismatic or basal slip system effects the rate and extent of the stress accumulation in the hard grain. The specifics of the so-called rogue grain pairs leading to the most extreme stress enhancement should perhaps be revisited in reassessing how texture and nearest neighbour micro-texture affect the probabilities of nucleating a crack in dwell fatigue loading.

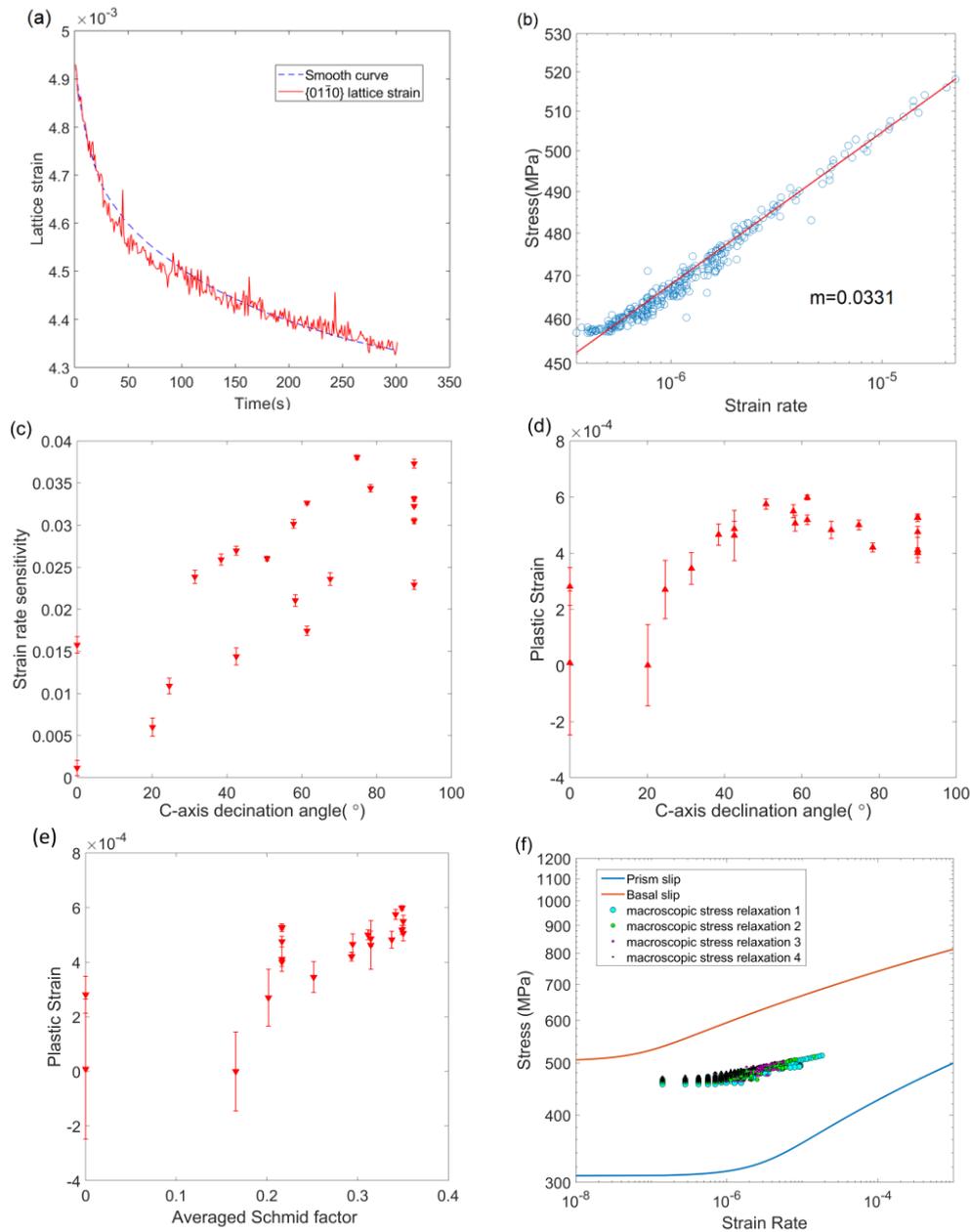

Figure 12: (a) Smoothing fit the first strain relaxation curve of the $\{01\bar{1}0\}$ plane; (b) Determination of strain rate sensitivity of $\{01\bar{1}0\}$ plane; (c) Strain rate sensitivity vs c-axis declination angle; (d) Plastic strain increment during the first macroscopic strain hold period for different grain families (diffraction rings) plotted as a function of c-axis declination angle; (e) Plastic strain increment during first

macroscopic strain hold period for different grain families (diffraction rings) plotted as a function of averaged Schmid factor of prism and basal slip. (f) Strain rate-stress variation given by the slip laws for prism and basal slip as well as macroscopic response measured by the ETMT during the first 4 periods of stress relaxation.

As shown in Figure 13, the change in the GND density during the first relaxation cycle were calculated from the model. For grains that accumulate a greater GND density during relaxation are generally in a 'soft' orientation (high c-axis declination angle) and have a higher averaged prism/basal Schmid factor. However, the scatter in the data is large and a clear trend is difficult to identify, implying that at grain-grain level, the plastic deformation not only depends on the orientation of a grain, but more importantly, the contribution of a neighbouring effect cannot be ignored. This observation corroborate well with other studies that show deformation in α–titanium is strongly affected by neighbouring grain orientations[50,51]. In the context of cold creep, load shedding was studied by assessing different combinations of the neighbouring grain orientations using the crystal plasticity model, a worst case scenario does exist in which a grain that is well oriented for slip has an active slip system oriented at about 30° to the direction of the adjacent hard grain c-axis[7]. This effect could be further understood with a suitable experiment dedicated to directly measuring GND densities during cold creep. Such data will direct modelling and aid calibration of crystal plasticity models. It should also be noted that these GND density changes are at most less than 20% of the initial mobile dislocation density value (which $\rho = 5 \ \mu m^{-2} = 5 \times 10^{12} \ m^{-2}$), and more typically less than 10%. This is in accord with the finding that a good fit could be established with the dislocation density held constant in the model, and the lack of any appreciable changes in the diffraction peak widths.

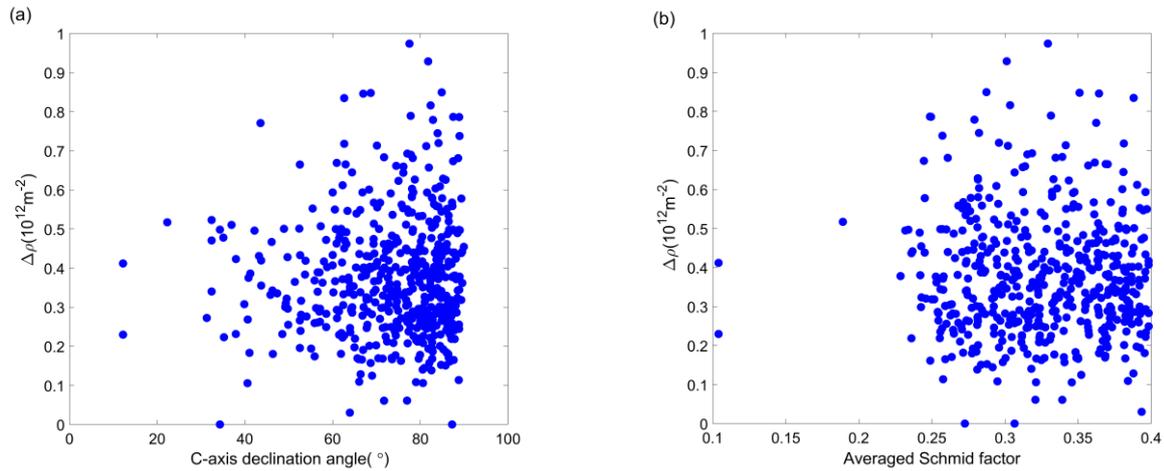

Figure 13. Simulated GND density increment during the first relaxation cycle for all the grains in the model vs. (a) c-axis declination angle of each grain and (b) averaged prism/basal Schmid factor of each grain.

## 4. Summary

An in-situ X-ray diffraction method was used to measure the lattice strain evolution in a commercially pure Ti grade 4 sample. The sample was deformed in tension to a small plastic strain, followed by a five-minute strain hold to allow stress relaxation to occur. The results were interpreted using a crystal plasticity finite element model. The study has the following key conclusions:

1. During loading, observation of individual lattice strains was used to identify the behaviour of different basal and prismatic slip systems. Combining with the crystal plasticity finite element analysis, key parameters in the slip rule for each slip system were isolated and determined.

2. Crystal plasticity modelling of lattice strain results were adjusted to match the experiment axial lattice strain results of $\{01\bar{1}0\}$, $\{01\bar{1}1\}$ and $\{11\bar{2}2\}$ plane families. The following parameters for basal and prismatic slip were found:

$\tau_c^{basal} = 252$ MPa, $\tau_c^{prism} = 154$ MPa, $\Delta F^{basal} = 10.5 \times 10^{-20}$ J (0.67eV) and $\Delta F^{prism} = 9.0 \times 10^{-20}$ J (0.56 eV).

3. The slip activation energy, $\Delta F$, and critical resolved shear stress, $\tau_c$, were found to be higher for basal slip over prismatic slip, indicating basal slip is more difficult to activate than prismatic slip. Coupled to this is a higher strain rate sensitivity for prism slip at the relatively low strain rates studies here.

4. The strain rate sensitivity was found to be higher in grains with a 'soft' orientation over those classified as 'hard'. The change in plastic strain during a stress relaxation period was also found to be higher in 'soft' grains.

## Data Availability

The synchrotron diffraction patterns and mechanical test data recorded during this experiment are openly available on the website https://zenodo.org/ as deposit 3960529 which can be found through the DOI: 10.5281/zenodo.3960529

## Author Contributions – CRediT (https://casrai.org/CRediT)

**Yi Xiong:** Data Curation, Formal Analysis, Methodology, Investigation, Software, Validation, Visualisation, Writing – Original Draft, Writing – Review & Editing

**Phani Karamched:** Data Curation, Formal Analysis, Investigation, Methodology, Software, Supervision, Validation, Writing – Review & Editing

**Chi-Toan Nguyen:** Investigation, Methodology, Validation, Writing – Review & Editing

**David M Collins:** Data Curation, Formal Analysis, Investigation, Methodology, Software, Writing – Review & Editing

**Christopher M Magazzeni:** Investigation, Writing – Review & Editing

**Edmund Tarleton:** Data Curation, Formal Analysis, Methodology, Software, Supervision, Validation, Writing – Review & Editing

**Angus J Wilkinson:** Conceptualization, Funding Acquisition, Investigation, Project Administration, Supervision, Visualisation, Writing – Review & Editing

## Acknowledgements

The authors acknowledge funding from the EPSRC through the HexMat programme grant (EP/K034332/1) and the Diamond Light Source for beam time under experiment EE17222. We are grateful for use of characterisation facilities within the David Cockayne Centre for Electron Microscopy, Department of Materials, University of Oxford, which has benefitted from financial support provided by the Henry Royce Institute (Grant ref EP/R010145/1). YX expresses gratitude to the financial support of China Scholarship Council (CSC) and ET acknowledges EPSRC for support through Fellowship grant (EP/N007239/1). We would like to thank Dr.Thomas Connolley, Dr.Robert Atwood and Dr.Stefan Michalik for their friendly and patient help at the beamline I12 and Dr.Nicolo Grilli for the helpful discussion.

## Appendix A. Evaluation of the effect of residual strain

The noise level of the lattice strain results were determined by fitting the experiment results with smooth curves (see Figure A.1). The Root-mean-squared deviation of the two curves was used to represent the noise level of the experimental result.

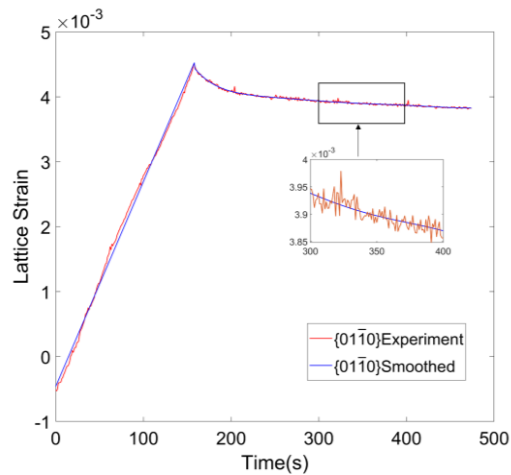

Figure A.1. Determination of noise level of experimental results of the $\{01\bar{1}0\}$ lattice strain.

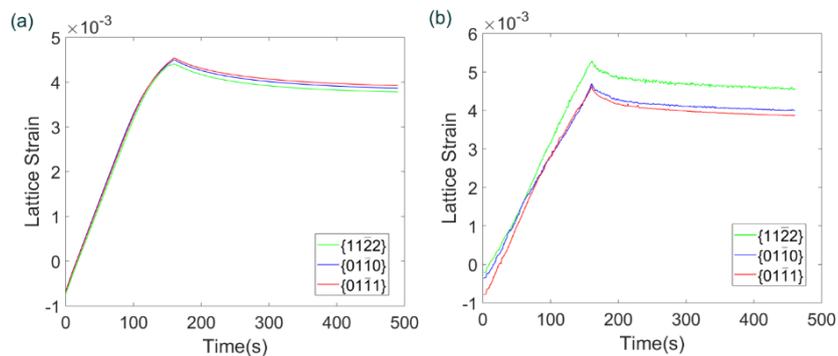

Figure A.2. (a) Modelled lattice strain with residual strain introduced by a quenching step; (b) Experimental lattice strain results.

The residual strain was introduced into the model by adding a quenching step (from $400°C$ to room temperature) prior to the mechanical loading step, as shown in Figure A.2a, the initial residual strain levels were in approximately the same range as the experiment (see Figure A.2b). There is little difference in the modelled residual strain in the $\{01\bar{1}0\}$, $\{01\bar{1}1\}$ and $\{11\bar{2}2\}$ planes, however, the experimental results show the difference among these three planes are much larger. It is possible that the residual strain in the sample was introduced not only by thermal but also by mechanical processes. The thermal expansion coefficient used in the model is stated in Table A.1.

Table A.1. Single crystal thermal expansion tensor coefficients for Ti[36,50]

| Coefficient | 11 | 22 | 33 |
|---|---|---|---|
| $\alpha$ | $1.8 \times 10^{-5}$ | $1.8 \times 10^{-5}$ | $1.1 \times 10^{-5}$ |

As the reflection-to-reflection variation was not so well captured by the model and the effects are in any case relatively small, we sought to simplify the analysis. To do this, we compared the change in lattice strain (the lattice strain in response to the external load) results from the model with introduced residual strain and a strain-free model running with the same material parameters. As shown in Figure A.3, the difference in lattice strain and macroscopic stress is small and less than the noise of the experimental result (see Table A.2). Therefore, the residual strain is considered to be negligible in this work.

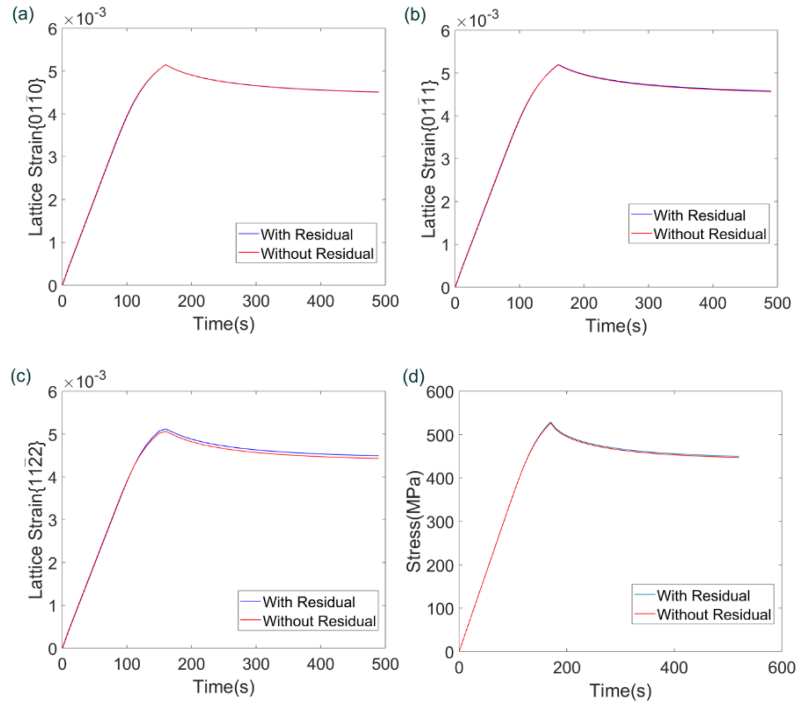

Figure A.3. Comparison of the model with and without residual strain of (a) $\{01\bar{1}0\}$ lattice strain; (b) $\{01\bar{1}1\}$ lattice strain; (c) $\{11\bar{2}2\}$ lattice strain; and (d) Macroscopic stress.

Table A.2. Comparison of the lattice strain difference caused by initial residual strain and noise of the experimental lattice strain results.

|  | $\{01\bar{1}0\}$ | $\{01\bar{1}1\}$ | $\{11\bar{2}2\}$ |
|---|---|---|---|
| Difference caused by residual strain | $8.1 \times 10^{-6}$ | $1.4 \times 10^{-5}$ | $4.9 \times 10^{-5}$ |
| Noise of experimental result | $4.9 \times 10^{-5}$ | $6.3 \times 10^{-5}$ | $7.4 \times 10^{-5}$ |

# Appendix B. Justification of element size using lattice strain evolution

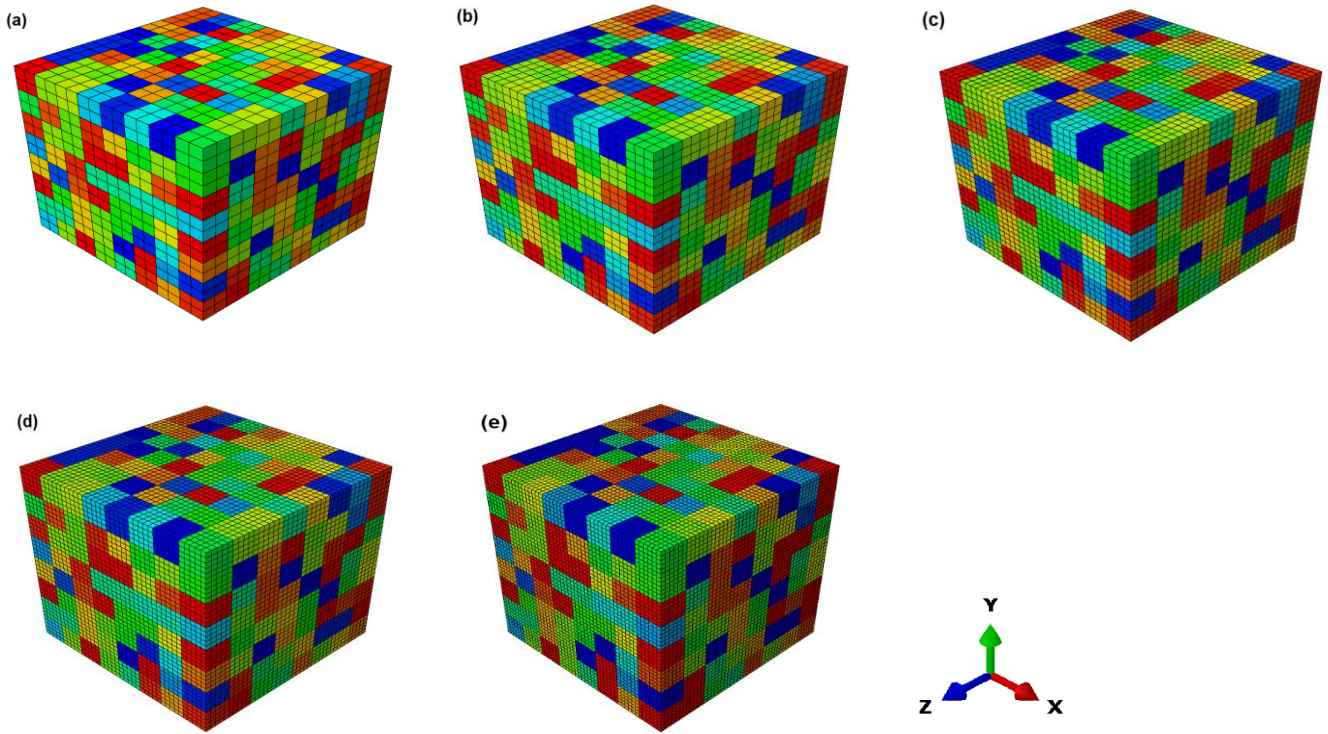

Figure B.1. The range of polycrystal models with differing number of elements within each grain (a) $2 \times 2 \times 2$ elements; (b) $3 \times 3 \times 3$ elements; (c) $4 \times 4 \times 4$ elements; (d) $5 \times 5 \times 5$ elements; (e) $6 \times 6 \times 6$ elements, which are used to justify the mesh size.

Five polycrystal models each with 512 grains shown in Figure B.1 were simulated with the same texture as the synchrotron stress relaxation test sample shown in Figure 1 and subjected to a uniaxial stress followed by a strain hold (same as the simulations shown in Figure 6b). Five models each had different mesh densities to justify the element size selected in the simulation. Lattice strain evolution of the $\{01\bar{1}0\}$ plane was compared with the finest mesh size model ($6 \times 6 \times 6$ elements per grain) as shown in Figure B.2 The results show that $2 \times 2 \times 2$ and $3 \times 3 \times 3$ elements per grain models show significant difference compared to the $6 \times 6 \times 6$ model, and the lattice strain relaxation curves for both of them are above that of the $6 \times 6 \times 6$ model. However, as the number of elements increased to $4 \times 4 \times 4$ and $5 \times 5 \times 5$ elements per grain, little change in the relaxation curves were observed. Although the differences between the simulation outputs are systematic, the root-mean-squared deviation of the lattice strain output from the five models were calculated using the $6 \times 6 \times 6$ model with the finest element size as the reference. Figure B.3 shows the deviation of the simulation results as a function of number of elements per grain. The red dashed line in Figure B.3 illustrates the mean error of the experiment results, and it is clear that for the $4 \times 4 \times 4$ elements per grain model, the mesh convergence issues were smaller than the experimental noise, and so this was assumed sufficient for simulations presented in this paper.

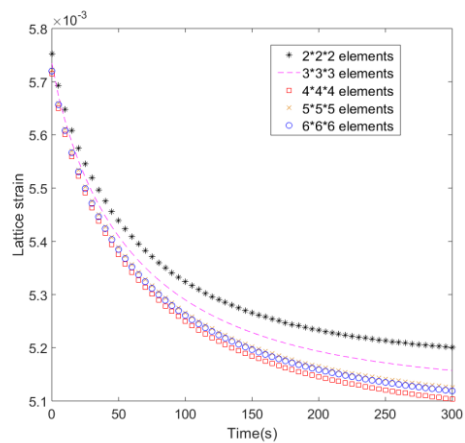

Figure B.2. The comparison of $\{01\bar{1}0\}$ plane lattice strain relaxation curves of the five models.

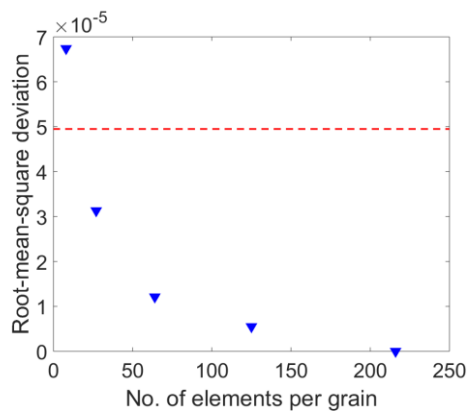

Figure B.3. Root-mean-squared deviation of the five models compared to the $6 \times 6 \times 6$ elements per grain model. The dashed red line indicates level of experimental noise.

# Appendix C. Sensitivity of lattice strain with changing parameters

Efforts were made to study the effect of changing the CRSS of pyramidal slip on the lattice strain relaxation curves, the results are shown in Table C.1. The root-mean-square deviation (RMSD) was calculated for the $\{01\bar{1}0\}$ lattice strain with CRSS for pyramidal slip set to be 2 times, 3 times and 4 times the value of basal slip, compared to the $\{01\bar{1}0\}$ lattice strain with pyramidal slip completely switched off. The results indicate that there was negligible activity of pyramidal slip in this experiment.

Table C.1. RMSD of the $\{01\bar{1}0\}$ lattice strain with different CRSS of pyramidal slip

| $\tau_c^{pyramidal}$ (MPa) | $2 \times \tau_c^{basal}$ | $3 \times \tau_c^{basal}$ | $4 \times \tau_c^{basal}$ |
|---|---|---|---|
| RMSD to no pyramidal slip | $1.03 \times 10^{-8}$ | 0 | 0 |

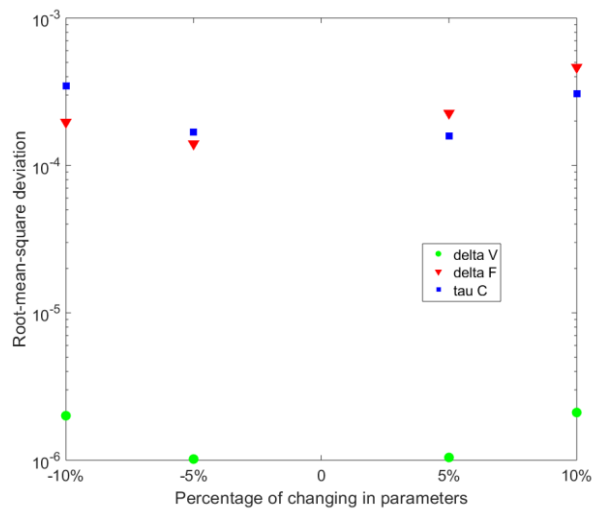

Figure C.1. Comparison of sensitivity of the lattice strain with respect to changing in parameters of ($\Delta V$, $\Delta F$ and $\tau_c$).

Figure C.1 shows the root-mean-square deviation between modelled lattice strain with changing a certain parameter and original modelled lattice strain. Compared to $\Delta F$ and $\tau_c$, $\Delta V$ results in a significantly less difference (about two order of magnitudes) in lattice strain.

Therefore, efforts were not been made to calibrate the $\Delta V$ and $\Delta V$. Instead, they were fixed to the value obtained from macroscopic stress relaxation for both basal and prismatic slip systems.

# Appendix D. Sensitivity of lattice strain to the initial orientation assignment

A new model was created with discrete orientations that were randomly assigned from the EBSD measurement (same method as discussed in section 2.3). The pole figures of this new model are shown in Figure D.1. Two models were subjected to the same boundary conditions and using the same set of materials parameters. Comparison of lattice strain evolution of the $\{01\bar{1}0\}$, $\{01\bar{1}1\}$ and $\{11\bar{2}2\}$ lattice planes during the first stress relaxation cycle in these two models is shown in Figure D.2. it can be seen that although the lattice strain results shifts by a small amount, the overall fitting remains reasonably good. The averaged root-mean-square deviation (RMSD) of the lattice strain of these three planes compared to the original model is $6.63 \times 10^{-5}$ which is slightly higher than the average experimental error (see Figure B.3). However, a fitting error equivalent to the original model was obtained, the parameters require only a small refinement (as shown in Figure C.1). Therefore, the simulated lattice strain is not strongly sensitive to the initial orientation assignment if the overall texture is generally the same.

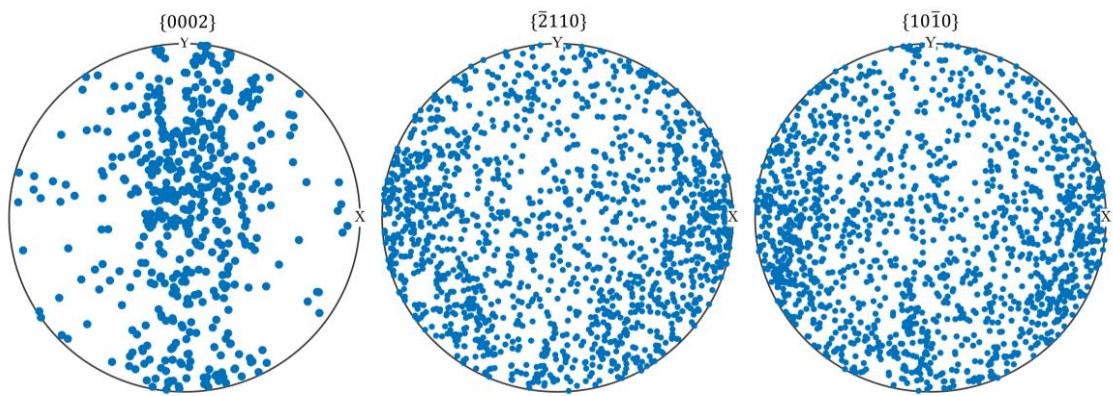

Figure D.1. Scatter pole figure plots of the new orientations assigned polycrystal finite element model.

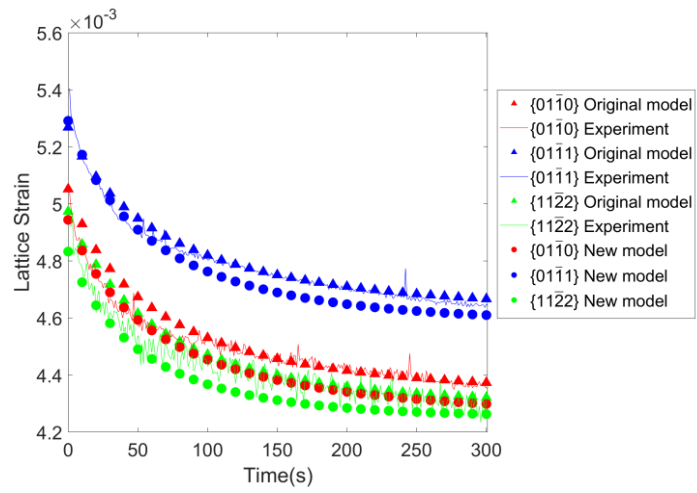

Figure D.2. Comparison of lattice strain evolution of $\{01\bar{1}0\}$, $\{01\bar{1}1\}$ and $\{11\bar{2}2\}$ planes during the first stress relaxation cycle between the original model and the new model.